\documentclass[letter]{aa}
\pdfoutput=1
\usepackage{graphicx}
\usepackage{txfonts}
\usepackage{multirow}
\usepackage{natbib}
\usepackage{color}
\usepackage{amsmath}
\newcommand{\gaia}{{\it Gaia~}}

\begin{document}

   \title{Origin of the system of globular clusters in the Milky Way}

   \subtitle{}

   \author{D. Massari\inst{1,2,3}
          \and
          H.H. Koppelman\inst{1}
          \and
          A. Helmi\inst{1}
          }

    \institute{Kapteyn Astronomical Institute, University of Groningen, NL-9747 AD Groningen, Netherlands
             \and
             Dipartimento di Fisica e Astronomia, Universit\`{a} degli Studi di Bologna, Via Gobetti 93/2, I-40129 Bologna, Italy 
	     \and
             INAF - Osservatorio di Astrofisica e Scienza dello Spazio di Bologna, Via Gobetti 93/3, I-40129 Bologna, Italy\\
              \email{davide.massari@unibo.it} 
              }

   \date{Received 19/06/2019; accepted 09/09/2019}


  \abstract
  {The assembly history experienced by the Milky Way is
    currently being unveiled thanks to the data provided by the {\it
      Gaia} mission.  It is likely that the globular cluster system of our Galaxy has
    followed a similarly intricate formation path.}
  {To constrain this formation path, we explore the link between the
    globular clusters and the known merging events that the
    Milky Way has experienced.}
  {To this end, we combined the kinematic information provided by
    {\it Gaia} for almost all Galactic clusters, with the largest
    sample of cluster ages available after carefully correcting for systematic errors.
    To identify clusters with a common origin we analysed their
    dynamical properties, particularly in the space of integrals of
    motion.}
  {We find that about 40\% of the clusters likely formed in
      situ. A similarly large fraction, 35\%, appear to be
      possibly associated to known merger events, in particular
    to {\it Gaia}-Enceladus (19\%), the Sagittarius dwarf galaxy
    (5\%), the progenitor of the Helmi streams (6\%), and to the
    Sequoia galaxy (5\%), although some uncertainty remains due to
    the degree of overlap in their dynamical characteristics. Of the
    remaining clusters, 16\% are tentatively associated to a group
    with high binding energy, while the rest are all on loosely bound
    orbits and likely have a more heterogeneous origin. The resulting
    age--metallicity relations are remarkably tight and differ in their
    detailed properties depending on the progenitor, providing further
    confidence on the associations made.}
  {We provide a table listing the likely associations. Improved
    kinematic data by future {\it Gaia} data releases and especially a
    larger, systematic error-free sample of cluster ages would help to
    further solidify our conclusions.}

   
   \maketitle
%

\section{Introduction}

According to the $\Lambda$CDM cosmological paradigm, structure
formation proceeds bottom-up, as small structures merge together to
build up the larger galaxies we observe today.  The Milky Way is a
prime example of this formation mechanism, as  first demonstrated by
the discovery of the Sagittarius dwarf spheroidal galaxy in the
process of disruption \citep{ibata94}, then in halo stellar streams crossing
the solar neighbourhood \citep{helmi99}, and more recently by the
discovery of stellar debris from Gaia-Enceladus, revealing the last significant merger experienced
by our Galaxy (\citealt{helmi18}, see
also \citealt{belo18}).

As a natural result of such events, not only field stars but also
globular clusters (GCs) may have been accreted
(\citealt{penarrubia09}).  Starting with \citet{searle78} there has
been a quest to understand which of the approximately 150 GCs hosted
by the Galaxy actually formed in situ and which formed in different
progenitors that were only later accreted.  Recently, the availability
of precise relative ages \citep[with formal errors of $\lesssim1$ Gyr;
e.g.][]{marin09, vandenberg13} and homogeneous metallicity
measurements (\citealt{carretta09}) led to the discovery that the
age--metallicity relation (AMR) of Galactic GCs is bifurcated
(\citealt{marin09, forbes10, leaman13}). Although limited, kinematic information
(e.g. \citealt{dinescu97, dinescu99, massari13}) nonetheless helped
to reveal that the metal-poor branch of young GCs have halo-like
kinematics (and are therefore more likely to be accreted), whereas GCs in the
young and metal-rich branch have disc-like kinematics, and are
consistent with having formed in situ (see also \citealt{recio18}).

With the advent of the second data release (DR2) of the {\it Gaia}
mission (\citealt{brown18}), we have, for the first time, full
six-dimensional phase space information for almost all of the Galactic
GCs \citep{gaiahelmi18,vasiliev19}.  Therefore, it is timely to
revisit the origin of the Galactic GC system. The goal of this {\it
  Letter} is to use this information to provide a more complete
picture of which GCs formed outside our Galaxy and in which progenitor
(among those currently known or yet to be discovered).  The main
result of this analysis is given in Table~\ref{tab1}, which lists all
the Galactic GCs and the progenitors they have likely been
  associated with. These associations may be seen to reflect the best
  of our current understanding, although some are only tentative. We
  may expect to be able to draw firmer conclusions with improved GC
  age datasets and larger samples of field stars with full-phase
  space kinematics.

\begin{figure*}
 \centering
    \includegraphics[width=15.5 cm, height=10.3 cm]{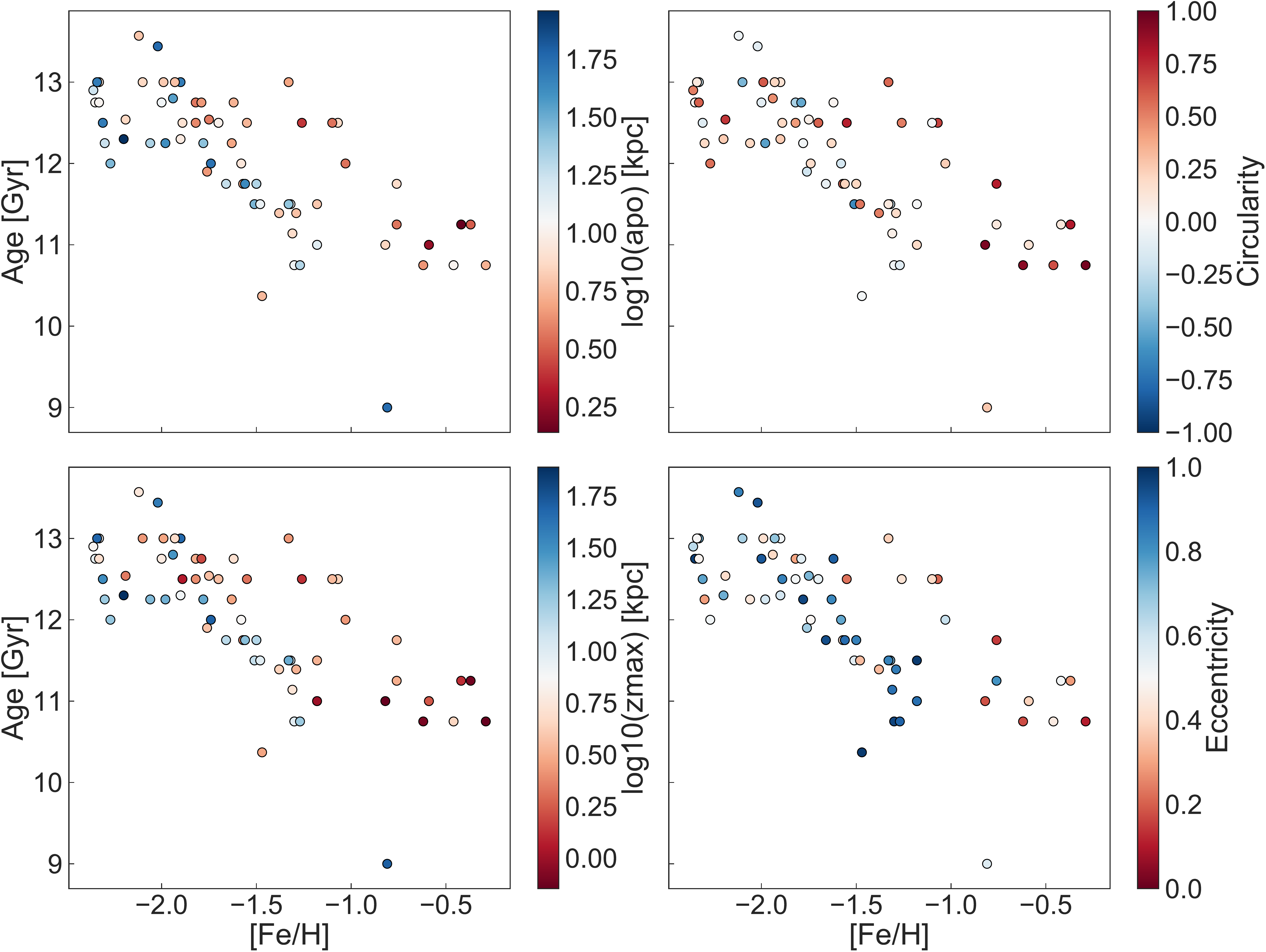}
        \caption{Age--metallicity relations for the sample of 69 GCs, colour-coded according to their dynamical properties.  We note that clusters on the young metal-rich branch share similar dynamical properties (in red), 
        and that clusters with these characteristics are also present for low metallicities and are typically older.}\label{amr_disk}
\end{figure*}

\section{The dataset: dynamical properties, ages, and metallicities}

We put together a dataset of $151$ clusters with full 6D
phase-space information of Galactic GCs known based on the
compilations by \cite{gaiahelmi18} and \cite{vasiliev19} (for
  more details, see Appendix \ref{sec:app-kin}). We used the
{\tt AGAMA} package (\citealt{vasiliev19b}) with the \cite{mcmillan17}
potential to compute GCs orbital parameters like the apocenter ({\sf
  apo}), maximum height from the disc ($Z_{\rm max}$), and eccentricity
({\sf ecc}).  We also computed the orbital circularity as {\sf
  circ} = $L_Z / L_{Z,{\rm circ}}$, where $L_{Z,{\rm circ}}$ is the
angular momentum of a circular orbit with the cluster energy, which thus
takes extreme values +1 or -1 for co-planar circular prograde or retrograde
orbits, respectively.

In this work we study the AMR for a subsample of GCs that has a
homogeneous set of ages and metallicities. This sample includes the
catalogue by \cite{vandenberg13}, who provide absolute ages and
uncertainty for 55 GCs, and objects from the compilation by
\cite{forbes10} who gathered relative age estimates from
\cite{salaris98}, \cite{deangeli05}, \cite{marin09}, which we add after checking for
systematic offsets with respect to \citet[]{vandenberg13}.  Our final
sample consists of 69~GCs with ages and metallicities (see
  Appendix \ref{sec:app-ages} for details and a discussion of the care
  needed when handling GC ages).

\section{Assignment of clusters}

Figure \ref{amr_disk} shows the AMR for the clusters in our sample,
colour-coded according to various dynamical properties, namely {\sf
  apo, circ}, $Z_{\rm max}$ and {\sf ecc}.  It is immediately clear
that the clusters located on the young and metal-rich branch of the
AMR are dynamically different from those on the young and metal-poor
branch.  Young and metal-rich GCs typically do not reach high
altitudes above the Galactic plane ($Z_{\rm max}$), have smaller
apocentres, and tend to have lower eccentricities. As already
recognised in the literature (e.g. \citealt{leaman13}), these are the
clusters formed { in situ}, either in the disc or in the bulge, in
what we hereafter refer to as the {Main Progenitor}. For the first time
we can recognise from Fig.~\ref{amr_disk} some old metal-poor GCs
with orbital properties characteristic of the young metal-rich
branch, which thus would also belong to the {Main Progenitor}.

\subsection{In situ clusters} 

 We therefore use the dynamical properties of the GCs that
  populate the young and metal-rich branch of the AMR of
  Fig.~\ref{amr_disk}, and which were likely born in the Milky Way,
to define simple criteria to identify {Main Progenitor} clusters: 
\begin{itemize}
\item Bulge clusters: those placed on highly bound orbits (with {\sf
    apo} < 3.5 kpc).  There are 36 GCs selected in this
  way\footnote{Nuclear clusters of dwarf galaxies accreted long ago
    could also end up on bulge-like orbits due to dynamical
    friction. The current uncertainty on the  proper motions of bulge GCs
    (they are typically highly extincted and only few stars are detected by
    {\it Gaia}), and the lack of an age estimate for most of them,
    prevents us from investigating this possibility.}.
\item Disc clusters satisfy: $i$) $Z_{\rm max}<5$ kpc and $ii$)
  ${\sf circ}>0.5$.  While this does not guarantee a ``pure''
    disc sample, and there may be a small amount of contamination, the
    effectiveness of these criteria is supported by the fact that the
  vast majority of the selected clusters tend to describe an AMR
  qualitatively similar to that found by \cite{leaman13}, except for
  two clusters located on the young and metal-poor branch (NGC6235 and
  NGC6254). The dynamical parameters of these  clusters are in the extremes of
  those characteristic of the Main Progenitor, and therefore we
  exclude them. This leaves a total of 26 Disc clusters.
\end{itemize}
We note that for [Fe/H]~$<-1.5$, { Main Progenitor} clusters are older
than the average. The $62$ { in situ} clusters are listed in
Table\ref{tab1} as Main-Disc (M-D) or Main-Bulge (M-B).

\begin{figure*}
    \includegraphics[width=\textwidth]{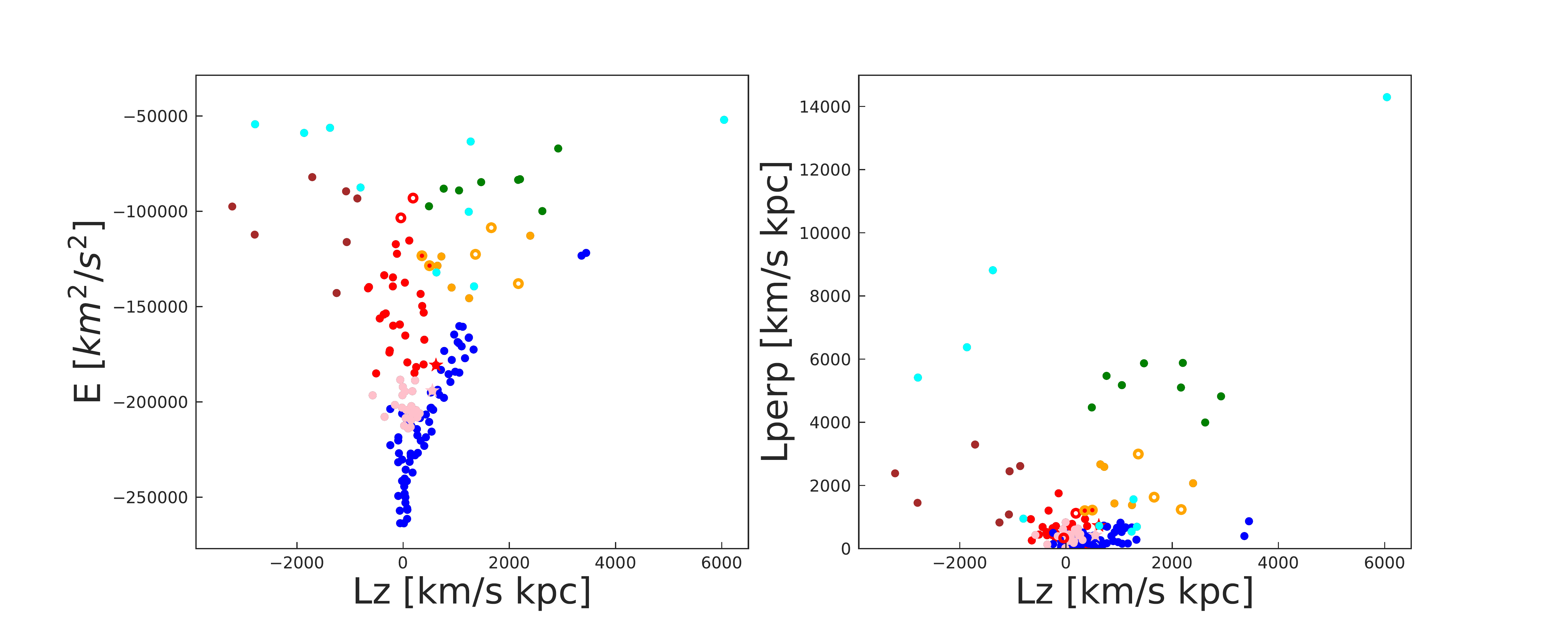}
    \caption{Two projections of IOM space for the 151 GCs in our
      sample, colour-coded according to their associations with
      different progenitors (blue symbols mark the {\it Main Progenitor},
      red is for {\it Gaia}-Enceladus, green for Sagittarius, orange for the 
      progenitor of the Helmi streams, brown for Sequoia, pink for the 
      low-energy group, and cyan for the high-energy group). 
      For visualisation purposes, two clusters
      (Crater and E1) with extremely negative $L_Z$ have not been
      plotted. Empty symbols correspond to tentative associations. The two
      star symbols mark the young and metal-poor clusters excluded by the 
      Disc selection.}\label{iom}
\end{figure*}

\subsection{Accreted clusters}

We now analyse the remaining clusters looking for a common association
with the progenitors of known merger events experienced by the Milky
Way. To do so, we investigate the integrals of motion (IOM) space
defined by $E$, $L_Z$, and $L_{perp}$, the latter being the angular
momentum component perpendicular to $L_Z$, which, despite not being
fully conserved in an axisymmetric potential like that of the Milky
Way, still helps in discriminating groups of stars (or clusters) with
similar origin (\citealt{helmi00}), also if the potential has
  varied with time \citep[see][]{penarrubia06, gomez13}. 
  In particular, we use the (known) extent
  of each progenitor stellar debris in the IOM space to provisionally
  identify associated GCs.

\subsubsection{The Sagittarius dwarf spheroidal galaxy}

The Sagittarius dwarf spheroidal galaxy constituted the first
discovery of a merger with the Galaxy \citep{ibata94}. By exploiting
numerical models that accurately reproduce the position and radial velocity
of stars belonging to the Sagittarius streams, \cite{lm10} provided a
list of candidate GCs that could have been associated to the
dwarf. This list has been recently refined by adding the information
on their proper motions as measured from HST and {\it Gaia}
observations (\citealt{massari17, sohn18}), and currently includes 6
GCs, namely M54, Arp~2, Pal~12, Terzan~7, Terzan~8, and NGC2419.

These six clusters describe a well-defined subgroup in IOM space:
$i$) $3700\!<\!L_{perp}\!<\!6200$~km/s~kpc and $ii$)
$0<L_Z<3000$~km/s~kpc, as seen in Fig.~\ref{iom}. Two more
clusters are found in this region of IOM space: NGC~5824 and
Whiting~1, which had previously been tentatively
associated with the dwarf by \cite{bellazzini03} and \cite{lm10},
respectively.

\subsubsection{The progenitor of the Helmi streams}

Recently, \cite{koppelman18} used {\it Gaia} data to characterise
the progenitor of the Helmi streams \citep[hereafter
H99,][]{helmi99}. According to these authors and based on their
dynamical properties (and a comparison with a numerical simulation
  that best reproduces the properties of the streams), the seven GCs
shown in orange in Fig.~\ref{iom} could be associated to this object.
Interestingly, these seven clusters were shown to follow a tight and
low-normalisation AMR.

To explore whether or not additional members could exist, we started from
the dynamical criteria suggested in their work, and revisited the
location of the selection boundaries while requiring consistency with
the AMR. The following criteria: $i$) $350<L_Z<3000$ km/s~kpc, $ii$)
$1000<L_{perp}<3200$ km/s~kpc, and $iii$)
$E <-1.0\times10^5$~km$^2$/s$^2$ seem to be the most
appropriate. Allowing for lower values of $L_Z$ leads to the inclusion
of very old clusters (not consistent with the AMR of the core members)
whereas increasing the upper limit leads to including two disc
clusters lacking an age estimate (Pal~1 and BH~176). Increasing the
$E$ limit would mean including a cluster with ${\sf apo}$ >100~kpc,
whereas the typical value for the core members is $\sim30$ kpc.
Finally, the limits on $L_{perp}$ are given by Sagittarius clusters on
one side and by the consistency with the AMR on the other.

Out of the ten GCs that would be associated to H99 according to
the above criteria, there is no age information for three of them: Rup~106, E3, and
Pal~5, and therefore we consider them to be tentative members (orange open
symbols in Fig.~\ref{iom}).  We note that Pal~5 and E3 have the lowest
{\sf ecc} ($\sim0.2$) in the set, with E3 having the lowest
$Z_{\rm max}$ ($\sim7$ kpc), and Rup~106 the largest {\sf apo}
($\sim34$~kpc).  Also in comparison to the H99 stars \citep[see][for
details]{koppelman18}, E3 and Pal~5 have more extreme orbital
properties, with Rup~106 being on a looser orbit. However, this does
not necessarily preclude membership since GCs are expected to be
less bound than the stars.

\subsubsection{{\it Gaia}-Enceladus}

\begin{figure}
    \includegraphics[width=\columnwidth]{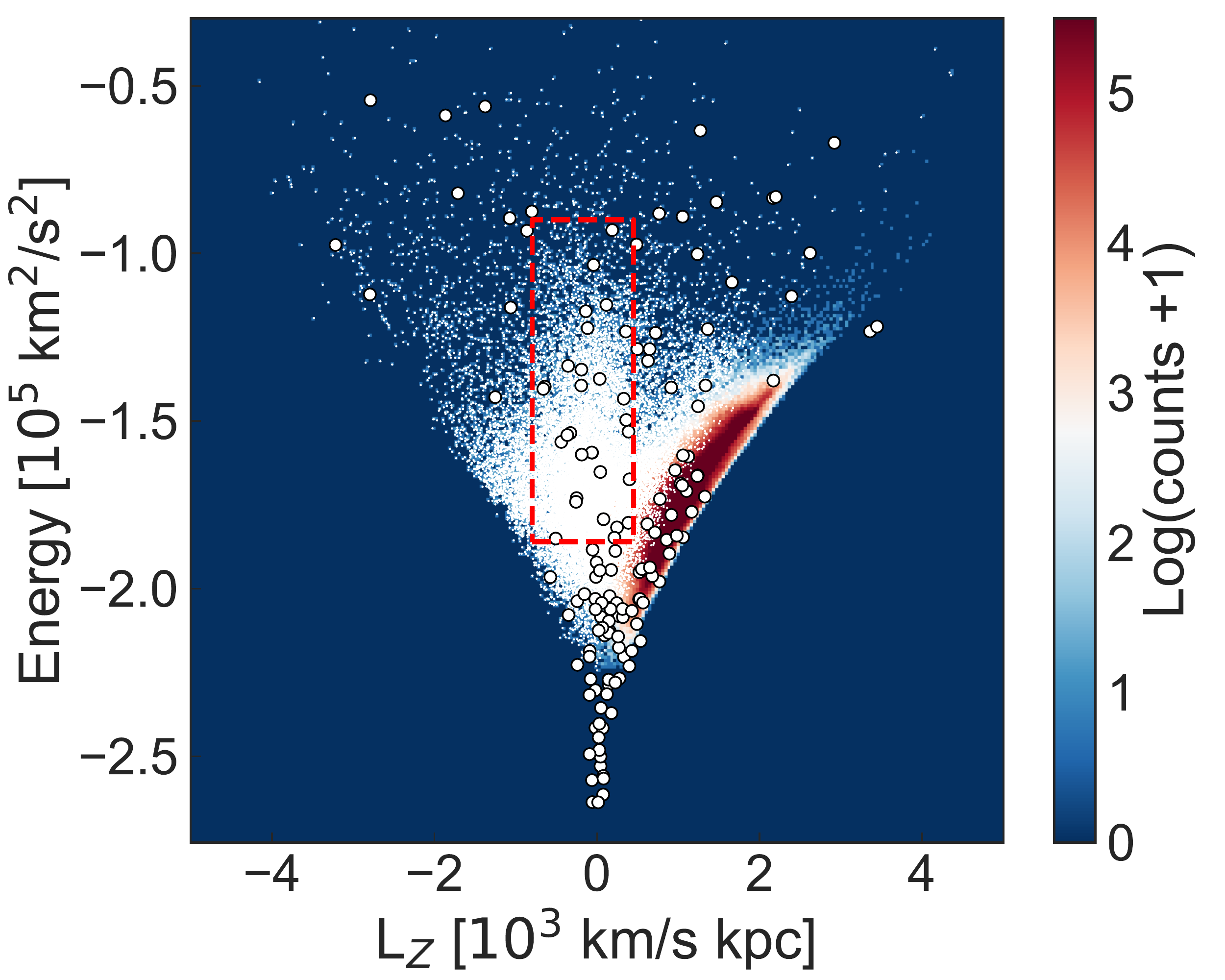}
    \caption{Integrals of motion for halo stars (white dots), Galactic disc stars
      (coloured contours), and our sample of GCs (white filled
      circles). The red box corresponds to the location of the stellar
      debris of G-E (\citealt{helmi18}), and is therefore used to
      select G-E clusters.}\label{ge_stars}
\end{figure}

To look for GCs associated to {\it Gaia}-Enceladus \citep[G-E
hereafter,][]{helmi18}, we directly compare the distribution in IOM
space of GCs to that of field stars with 6D kinematics from {\it
  Gaia}, as shown in Fig.~\ref{ge_stars}.

Based on this comparison, we associate clusters to G-E according to
the following criteria: $i$) $-800<L_Z<620$~km/s~kpc, $ii$)
$-1.86\times10^5<E<-0.9\times10^5$~km$^2$/s$^2$, and $iii$)
$L_{perp}<3500$~km/s~kpc.  This selection associates 28 GCs to
G-E. With the exception of NGC~7492 (its {\sf apo} being $\sim28$ kpc),
all of them have apocenter values $<25$ kpc, as reported for 
G-E stars \citep[see][]{deason18}.
The resulting AMR is remarkably tight (see below), and this
tightness can be used to explore the energy boundaries. By decreasing
the lower limit of $E$, a very old globular cluster enters the
selection which is significantly off the AMR described by the other
members. This energy limit is slightly higher than that used by
  \citet[who adopt the same Galactic potential as ours]{myeong18} to
  define the clusters belonging to the progenitor of the same
  accretion event, that they call Gaia-Sausage (see also
  \citealt{belo18}).  Moving the upper limit to
$E =-1.1\times10^5$~km$^2$/s$^2$ excludes Pal~2 and Pal~15, which we
therefore consider to be tentative members (open symbols in
Fig.~\ref{iom}).

Some of the associated clusters are located in regions of the IOM
  space that are shared by stellar debris of different progenitors.
  Therefore, the associations are more uncertain and deserve further
  discussion \citep[see e.g.][]{borsato19}. 
  For example, two clusters (NGC5904 and NGC5634) lie near
the region occupied by H99 debris. They have {\sf ecc} $\sim0.8$, and
their $Z_{\rm max}$ and {\sf apo} are somewhat larger but not
inconsistent with those typical of G-E clusters. While NGC5634 has no
age estimate, the age of NGC5904 ($11.5$ Gyr old) is consistent with both
the AMRs. We therefore consider them as tentative members to both
progenitors.
 
Also of particular interest is the overdensity of stars seen in
Fig.~\ref{ge_stars} at $L_Z\sim-3000$ km/s~kpc and $E \sim-10^5$
km$^2$/s$^2$, which has been associated to G-E by \cite{helmi18}
because of its resemblance to a feature seen in numerical simulations
of a merger event with similar characteristics to G-E
(\citealt{villalobos08}).  Two GCs (NGC3201 and NGC6101) are located
in this region of IOM space, and therefore we mark them as tentative
members but discuss them further in the following section.

With our selection criteria, $\omega$-Cen\footnote{for
  which we used the metallicity of the most metal-poor and oldest
  population, see \citealt{bellini17}.} is the cluster with
the highest binding energy among those associated to G-E. This is in good agreement with the idea that this cluster is in reality the remnant of the
nuclear star cluster of an accreted dwarf (e.g. \citealt{bekki03}) as
suggested by its peculiar chemistry.

After these considerations, we are left with $26$ possible members of
G-E (and $6$ tentative ones). Although this number is large, it is
consistent within the scatter with the relation between the number of
GCs and host halo mass \citep{vandokkum17}, given the mass estimate of
$6\times10^{10}$ M$_{\odot}$ from \cite{helmi18}.

\subsubsection{\it Sequoia}\label{seq}

Recently, \cite{myeong19} proposed the existence of merger debris
from a galaxy named {\it Sequoia} that would have been accreted
about 9 Gyr ago. Based on clustering algorithms performed over a
scaled action space, these authors found five GCs likely associated to
this system.  We find seven GCs to be possibly associated when considering
a selection box in $E$ and $L_{Z}$ corresponding to the stars from
{\it Sequoia} according to \cite{myeong19}, namely
$-3700<L_{Z}<-850$~km/s~kpc and
$-1.5\times10^5<E<-0.7\times10^5$~km$^2$/s$^2$.  Three of these GCs are in
common, namely FSR~1758, NGC~3201, and NGC~6101.  The other four
(IC~4499, NGC~5466, NGC~7006 and Pal~13) were excluded by Myeong et
al.  because of their slightly larger eccentricity ($\langle${\sf
  ecc}$\rangle \sim 0.75$, compared to their initial estimate of
$\sim0.6$).  Three clusters have known ages and these follow a low-normalisation AMR similar to the H99 GCs, which is consistent with the
low stellar mass estimated for {\it Sequoia} (\citealt{myeong19}).

As mentioned earlier, the {\it Sequioa} IOM selection has some overlap
with the arch-like overdensity ascribed to G-E debris by
\citet{helmi18}.  Discerning which is the actual progenitor of NGC3201
and NGC6101 is therefore not possible, and so in Table~\ref{tab1} we
link them to both systems. Nonetheless, there may be a slight
preference for {\it Sequoia} given their ages and metallicities. On
the other hand, \cite{myeong19} associated $\omega$-Cen and
NGC6535 to {\it Sequoia} because of their location in IOM
space. However, these two GCs follow a much higher AMR, typical of
clusters from more massive progenitors.  For this reason, we
prioritise the association of $\omega$-Cen to G-E and of NGC6535 to
one of the groups described below, though we acknowledge both the
  interpretations in Table~\ref{tab1}.

\subsubsection{The remaining clusters}\label{unassociated}

We were not able to associate 36 of the 151 GCs with full
phase-space information to known merger events. From their
distribution in the IOM space (Fig.~\ref{iom}), it is clear that at
least a significant fraction of them (25) could tentatively be part of a
  structure at low energy, with $ E<-1.86\times10^5$~km$^2$/s$^2$,
low $L_{perp}$, and with $L_{Z}\sim0$~km/s~kpc (pink symbols in
Fig.~\ref{iom}); we label these L-$E$.

The remaining 11 GCs all have high energy ($E >-1.5\times10^5$
km$^2$/s$^2$, in cyan in Fig.~\ref{iom}), but span a very large range
in $L_{Z}$ and $L_{perp}$.  Therefore, they cannot have a common
origin. Most likely instead, they have been accreted from different
low-mass progenitors which have not contributed debris (field stars)
to the Solar vicinity (as otherwise we would have identified
corresponding overdensities in Fig.~\ref{ge_stars}).  For convenience
only, we use a single label for all these objects (H-$E$, for
high energy) in Table~1.  Upcoming datasets, especially of field stars
with full phase-space information across the Galaxy, could be key to
understanding their origin given their heterogeneous properties.

\begin{figure}
    \includegraphics[width=8.5cm,trim=3cm 3cm 5cm 5cm,clip]{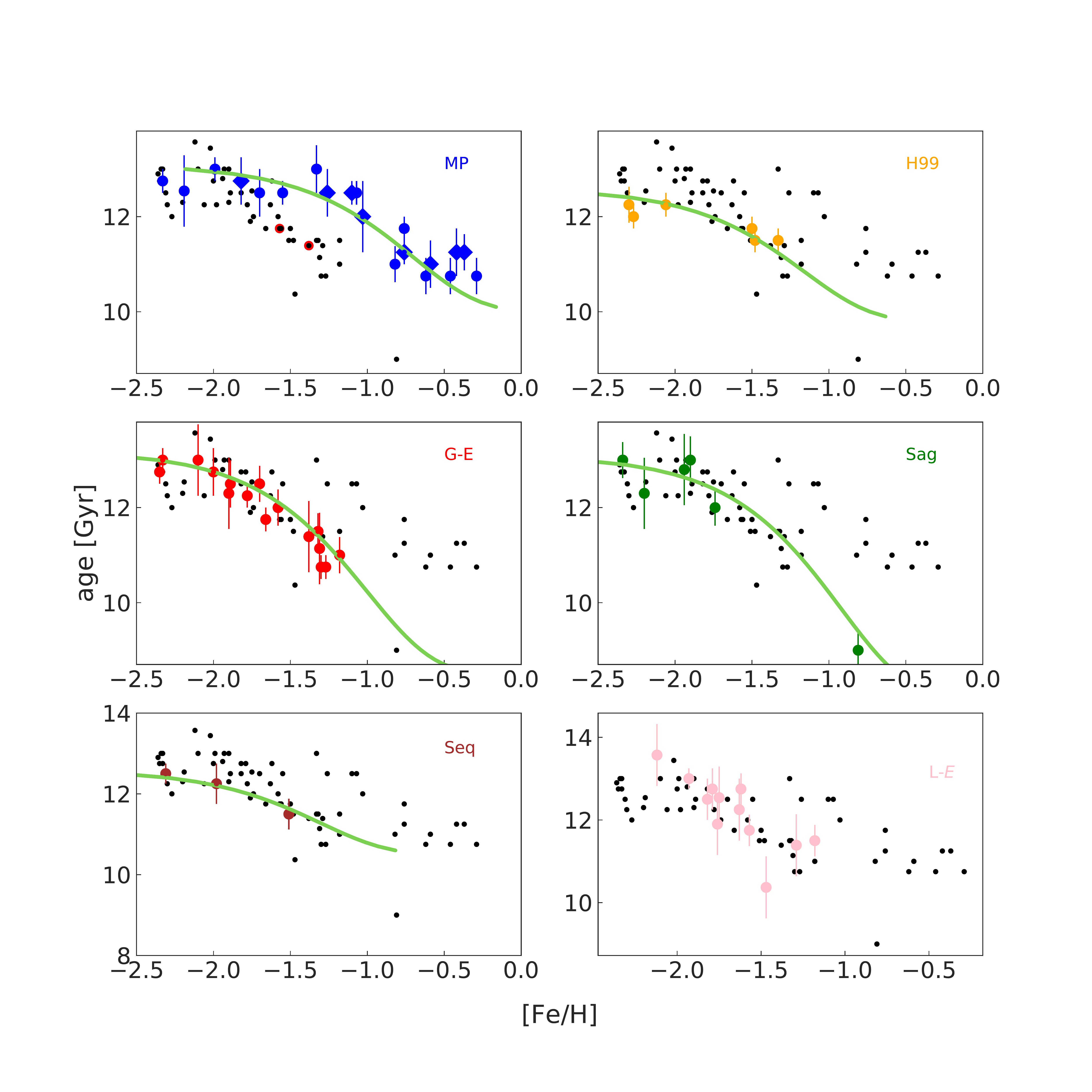}
    \caption{AMRs for the 69 GCs with age estimates, colour-coded as
      in Fig.\ref{iom}. The corresponding progenitors are marked in
      the labels. Individual age uncertainties are also plotted.
      In the AMR for {\it Main Progenitor} clusters (upper-left
      panel), diamonds represents bulge clusters while circles
      describe disc clusters.  The two red-circled black symbols are
      the two clusters that satisfy the disc membership criteria but
      that are excluded because they are near the boundary of the
      respective IOM region and their location in the AMR.}\label{amr}
\end{figure}

\subsection{Age--metallicity relation}\label{sect:amr}

The various panels in Fig.~\ref{amr} show the AMR of the clusters
associated to the different structures discussed so far, colour-coded
as in Fig.~\ref{iom}. Although consistency with the AMR of each
  group is checked after the IOM selection, it is quite remarkable
  that the dynamical identification of associations of GC results in
  AMRs that are all well-defined and depict different
  shapes or amplitudes.

The clusters of the  Main progenitor constitute the largest group
and have the highest normalisation, that is the most metal-poor oldest
clusters were born in the Galaxy itself.  The G-E AMR is remarkably
tight and has a high normalisation, though as expected this is not as
high as that of the { Main progenitor}. Similarly, the L-$E$ group
depicts a reasonably coherent AMR, with a high normalisation, which
seems even higher than that of G-E clusters, thus possibly suggesting
the presence of yet-to-be-discovered merger debris located
preferentially in the Galactic bulge and originating in a more massive
object. On the other hand, the AMR of the H99 members is remarkably low
in terms of normalisation, and this is consistent with the fact that
this progenitor is less massive ($M_\star \sim10^{8}$~M$_{\odot}$,
\citealt{koppelman18}).

We can describe the various AMRs with a leaky-box chemical evolution
model \citep{prantzos2008,leaman13,boecker19}, where the metallicity of the system evolves as
\begin{equation}
Z(t) = -p \ln \mu(t) = -p \ln {\frac{t_{f}-t}{t_f-t_i}} \;\;\; {\rm for } \; t \ge t_{i},
\end{equation}
where $\mu(t) = M_g(t)/M_g(0)$ is the gas fraction, and $p$ the
(effective) yield. We obtained this expression by assuming a
constant star formation rate starting at time $t_i$ after the Big Bang
and ending at time $t_f$, which we take to be the time of accretion
(which is constrained by estimates in the literature and which we took
to range from 3.2 Gyr for {\it Sequioa} to 5.7 Gyr for Sagittarius).
For the yield $p$ we assumed the dependence on $M_\star$ derived by
\citet{dekel2003} for star-forming dwarf galaxies \citep[see
also][]{prantzos2008}. The time $t_i$ is a free parameter which we
varied for each progenitor to obtain a reasonable description of the
observed points.  The only constraint we apply is that more massive
progenitors should start forming stars earlier, which led to $t_i$
values in the range of 0.5 Gyr (for the { Main progenitor}) to 1.1
Gyr (for {\it Sequioa}).  The resulting curves for each progenitor are
shown in the panels of Fig.~\ref{amr} with green solid lines. We
stress that these curves do not represent fits, but merely show that a
simple leaky box chemical evolution model is reasonably adequate to
describe the AMRs found for each set of clusters
associated to the individual progenitors.  The scatter around
  each curve (typically $<0.5$ Gyr) is qualitatively consistent with
  the individual age uncertainties. This might indicate that for the
  clusters lacking an age uncertainty, our assumed value of $0.75$ Gyr
  (see Appendix~\ref{sec:app-ages}) was too conservative, especially
  when considering relative ages rather than absolute ones.

\section{Summary and Conclusions}

In this Letter we exploit the complete kinematic information for 151
Galactic GCs, in combination with metallicity and homogeneous age
estimates for a subset of 69 GCs. Our goal is to elucidate which
GCs formed{ in situ} and which could have been accreted,
associating the latter to a particular progenitor based on their
dynamical properties and, where needed, on the shape of the AMR.

We found that 62 GCs likely formed in the Milky Way, and we separated
them into disc and bulge clusters based on their orbital parameters.
Among the accreted clusters, we assessed their possible associations
with the progenitor of four known merger events: {\it Gaia}-Enceladus,
the Sagittarius dwarf, the Helmi streams, and the {\it Sequoia} galaxy.
We identified $ 26$ (and an additional 6 tentative) GCs associated
to {\it Gaia}-Enceladus. This large number as well as the high
normalisation of their AMRs are consistent with {\it Gaia}-Enceladus
being the most massive among these four objects. According to our
findings, $\omega$-Centauri would be its nuclear star cluster.  We
further identified eight clusters associated with the Sagittarius dwarf,
possibly ten clusters to the progenitor of the Helmi streams, and a
plausible seven to {\it Sequoia}. Despite not being very populated, the
AMRs of these two groups are consistent with the lower literature
estimates for their mass. There is an inherent uncertainty to
  these assignments, because debris from different progenitors partly
  overlaps in IOM space, as in the case of {\it Sequoia} and G-E, and
  to a lesser degree for G-E and the Helmi streams.

  Interestingly, the specific frequency per unit of galaxy mass
  (T$_N$) as defined by \cite{zepf93} for each progenitor closely follows the relation reported in \cite{zaritsky16} for a sample of galaxies
  in the same stellar mass range. For example, we find $\log T_N$ =
  [1.8, 2.0, 1.2, 2.1] for G-E, the Helmi streams, Sgr, and {\it
    Sequioa}, while the expected values using \cite{zaritsky16} are
  respectively, [1.8, 2.2, 1.8, 2.4], which is well within the
  observed scatter of the relation ($\sigma(\log T_N)\sim0.7$).
  The values of $\log T_N$ were computed using stellar mass estimates of  
  [5, 1, 5]$\times10^{8} M_{\odot}$ for G-E, the Helmi streams, and Sagittarius. 
  These were derived with different methods, using chemistry in the first 
  case (\citealt{helmi18}) and N-body models for the latter two 
  (\citealt{koppelman18, dier17}). In the case of Sequioa, we used the mass 
  estimated by \cite{myeong19} on the basis of the specific frequency, 
  and it is therefore less surprising to find good agreement 
  (although our estimate uses a different number of associated clusters).

The 36 clusters that we have not associated to known debris
can be split in two groups based on orbital energy.  While the class
of GCs with low binding energy is very heterogeneous and likely has
several sites of origin, the low-energy (highly bound) group with 25
tentative members is relatively highly clustered in its dynamical properties and
shows a reasonably tight and high-normalisation AMR, possibly
suggesting the presence of debris towards the Galactic bulge from a  large
hitherto unknown galaxy. This finding is consistent with
  the conclusion that another significant accretion event is required
  to explain the overall AMR of Galactic GCs; for example, a merger with a
  ``Kraken''-like galaxy \citep{kruijssen18}. We note however, that most
  of the clusters reported in \citet{kruijssen18} as possible members of Kraken
  are not dynamically coherent, since three are in common with the H99
  GCs, seven with G-E, two with Sequoia, and one with the Main Progenitor.
  Therefore, although the L-$E$ group is Kraken-like, the associated
  clusters are different. It transpires that taking into account the
  dynamical properties is fundamental to establishing the
  origin of the different GCs of our Galaxy.

The next data release of the {\it Gaia} mission will provide improved
astrometry and photometry for all the GCs, as well as for a much larger
sample halo stars.  This will be crucial to achieving a complete and
accurate sample of GCs with absolute ages. Moreover it will lead to a
better understanding of the debris of the known progenitors, and
possibly to the discovery of new ones. The combination of these
factors will result in significant progress in the field and allow to
better pin down tentative associations and possibly also to assess
under which conditions the different clusters formed, such as for
example whether formation was prior to or during the different merger events.

\begin{acknowledgements}
We thank the anonymous referee for comments and suggestions
which improved the quality of our paper.
DM, HHK and AH acknowledge financial support from a Vici grant from NWO.
This work has made use of data from the European Space Agency (ESA)
mission \gaia (http://www.cosmos.esa.int/gaia), processed by the
\gaia Data Processing and Analysis Consortium (DPAC, {\tt
http://www.cosmos.esa.int/web/gaia/dpac/consortium}). Funding for
the DPAC has been provided by national institutions, in particular
the institutions participating in the \gaia Multilateral Agreement.

\end{acknowledgements}

\begin{appendix}
 
\begin{table*}
\centering
\caption{List of Galactic GCs and associated progenitors. M-D stands for Main-Disc; M-B stands for Main-Bulge;
G-E stands for {\it Gaia}-Enceladus; Sag stands for Sagittarius dwarf; H99 stands for Helmi Streams; Seq stands for Sequoia galaxy; 
L-E stands for unassociated Low-Energy; H-E stands for unassociated High-Energy. Finally XXX mark clusters with no available 
kinematics. Most of these are bulge GCs too extincted to be observed by Gaia.}\label{tab1}
\begin{tabular}{cccccccc}
\hline
GC & Progenitor  &   GC   &  Progenitor   & GC & Progenitor & GC & Progenitor \\
   &             &        &               &    &            &    &  \\
\hline
   NGC~104    & M-D  &  NGC~5927 & M-D &  HP 1     & M-B  &  GLIMPSE~2& XXX  \\
   NGC~288    & G-E  &  NGC~5946 & L-E &  NGC 6362 & M-D  &  NGC~6584& H-E  \\
   NGC~362    & G-E  &  BH~176   & M-D &  Liller 1 & XXX  &  NGC~6624& M-B  \\
   Whiting~1  & Sag  &  NGC~5986 & L-E &  NGC 6380 & M-B  &  NGC~6626& M-B  \\
   NGC~1261   & G-E  &  Lynga~7  & M-D &  Terzan 1 & M-B  &  NGC~6638& M-B  \\
   Pal~1      & M-D  &  Pal~14   & H-E &  Ton 2    & L-E  &  NGC~6637& M-B  \\
   AM~1       & H-E  &  NGC~6093 & L-E &  NGC 6388 & M-B  &  NGC~6642& M-B  \\
   Eridanus   & H-E  &  NGC~6121 & L-E &  NGC 6402 & L-E  &  NGC~6652& M-B  \\
   Pal~2      & G-E? &  NGC~6101 & Seq/G-E &  NGC 6401 & L-E  &  NGC~6656& M-D  \\
   NGC~1851   & G-E  &  NGC~6144 & L-E &  NGC 6397 & M-D  &  Pal~8   & M-D  \\ 
   NGC~1904   & G-E  &  NGC~6139 & L-E &  Pal 6    & L-E  &  NGC~6681& L-E  \\
   NGC~2298   & G-E  &  Terzan~3 & M-D &  NGC 6426 & H-E  &  GLIMPSE~1& XXX  \\
   NGC~2419   & Sag  &  NGC~6171 & M-B &  Djorg 1  & G-E  &  NGC~6712& L-E  \\
   Ko~2       & XXX  &  1636-283 & M-B &  Terzan 5 & M-B  &  NGC~6715& Sag  \\
   Pyxis      & H-E  &  NGC~6205 & G-E &  NGC 6440 & M-B  &  NGC~6717& M-B  \\
   NGC~2808   & G-E  &  NGC~6229 & G-E &  NGC 6441 & L-E  &  NGC~6723& M-B  \\
   E~3        & H99? &  NGC~6218 & M-D &  Terzan 6 & M-B  &  NGC~6749& M-D  \\
   Pal~3      & H-E  &  FSR~1735 & L-E &  NGC 6453 & L-E  &  NGC~6752& M-D  \\
   NGC~3201   & Seq/G-E &  NGC~6235 & G-E &  UKS 1    & XXX  &  NGC~6760& M-D  \\
   Pal~4      & H-E  &  NGC~6254 & L-E &  NGC 6496 & M-D  &  NGC~6779& G-E \\
   Ko~1       & XXX  &  NGC~6256 & L-E &  Terzan 9 & M-B  &  Terzan~7& Sag  \\
   NGC~4147   & G-E  &  Pal~15   & G-E?&  Djorg 2  & M-B  &  Pal~10  & M-D  \\
   NGC~4372   & M-D  &  NGC~6266 & M-B &  NGC 6517 & L-E  &  Arp~2   & Sag  \\
   Rup~106    & H99? &  NGC~6273 & L-E &  Terzan10 & G-E  &  NGC~6809& L-E  \\
   NGC~4590   & H99  &  NGC~6284 & G-E &  NGC 6522 & M-B  &  Terzan~8& Sag  \\
   NGC~4833   & G-E  &  NGC~6287 & L-E &  NGC 6535 & L-E/Seq  &  Pal 11  & M-D  \\
   NGC~5024   & H99  &  NGC~6293 & M-B &  NGC 6528 & M-B  &  NGC~6838& M-D  \\
   NGC~5053   & H99  &  NGC~6304 & M-B &  NGC 6539 & M-B  &  NGC~6864& G-E  \\
   NGC~5139   & G-E/Seq  & NGC~6316 & M-B &  NGC 6540 & M-B  &  NGC~6934& H-E  \\
   NGC~5272   & H99  &  NGC~6341 & G-E &  NGC 6544 & L-E  &  NGC~6981& H99  \\
   NGC~5286   & G-E  &  NGC~6325 & M-B &  NGC 6541 & L-E  &  NGC~7006& Seq  \\
   AM~4       & XXX  &  NGC~6333 & L-E &  2MS-GC01 & XXX  &  NGC~7078& M-D  \\
   NGC~5466   & Seq  &  NGC~6342 & M-B &  ESO-SC06 & G-E  &  NGC~7089& G-E  \\
   NGC~5634   & H99/G-E  &  NGC~6356 & M-D &  NGC 6553 & M-B  &  NGC~7099& G-E  \\
   NGC~5694   & H-E  &  NGC~6355 & M-B &  2MS-GC02 & XXX  &  Pal~12  & Sag  \\
   IC~4499    & Seq  &  NGC~6352 & M-D &  NGC 6558 & M-B  &  Pal~13  & Seq  \\
   NGC~5824   & Sag  &  IC~1257  & G-E &  IC 1276  & M-D  &  NGC~7492& G-E  \\
   Pal~5      & H99? &  Terzan~2 & M-B &  Terzan12 & M-D  &  Crater  & H-E  \\
   NGC~5897   & G-E  &  NGC~6366 & M-D &  NGC 6569 & M-B  &  FSR~1716& M-D  \\
   NGC~5904   & H99/G-E  &  Terzan~4 & M-B &  BH 261   & M-B  &  FSR~1758 & Seq  \\
\hline                                
\end{tabular}
\end{table*}

\section{Details of the sample of globular clusters}
\subsection{Kinematical properties}
\label{sec:app-kin}

To put together the sample of GCs, we started from the 75 GCs analysed
by \cite{gaiahelmi18}, who combined the {\it Gaia} measured proper
motions with distances and line-of-sight velocities available from the
compilation by \citet[][2010 edition]{harris96}. We then added the
data for the remaining GCs from \cite{vasiliev19}, who determined the
6D coordinates combining {\it Gaia} measurements with line-of-sight
velocities also from \citet{baum18}.

We then transformed the observed measurements of the kinematics of the
resulting catalogue of $151$ clusters to the Galactocentric reference
frame.  To this end we assumed a Local Standard of Rest velocity
$V_{\rm LSR} = 232.8$ km/s \citep{mcmillan17}, a solar motion
$(U,V,W) = (11.1, 12.24,7.25)$ km/s \citep{schonrich10}, and a distance
from the Sun to the Galactic centre of $R_0 = 8.2$ \citep{mcmillan17}.

\subsection{Homogeneous cluster ages}
\label{sec:app-ages}

Many methods have been applied to determine the
absolute age of a GC.  Photometric errors, poor calibration, and
uncertainties on the cluster distance and reddening however, can affect
such age estimates.  To overcome these, it has often been
preferred to determine relative ages (e.g. \citealt{buonanno89,
  bono10, massari16}), although these then need to be calibrated to
some absolute scale (e.g. \citealt{marin09}).  This explains why
available age compilations of GCs in the literature are so
heterogeneous, and therefore dangerous to blindly combine
together as different methods can result in systematic differences
that amount to several gigayears.

\begin{figure}
    \includegraphics[width=\columnwidth]{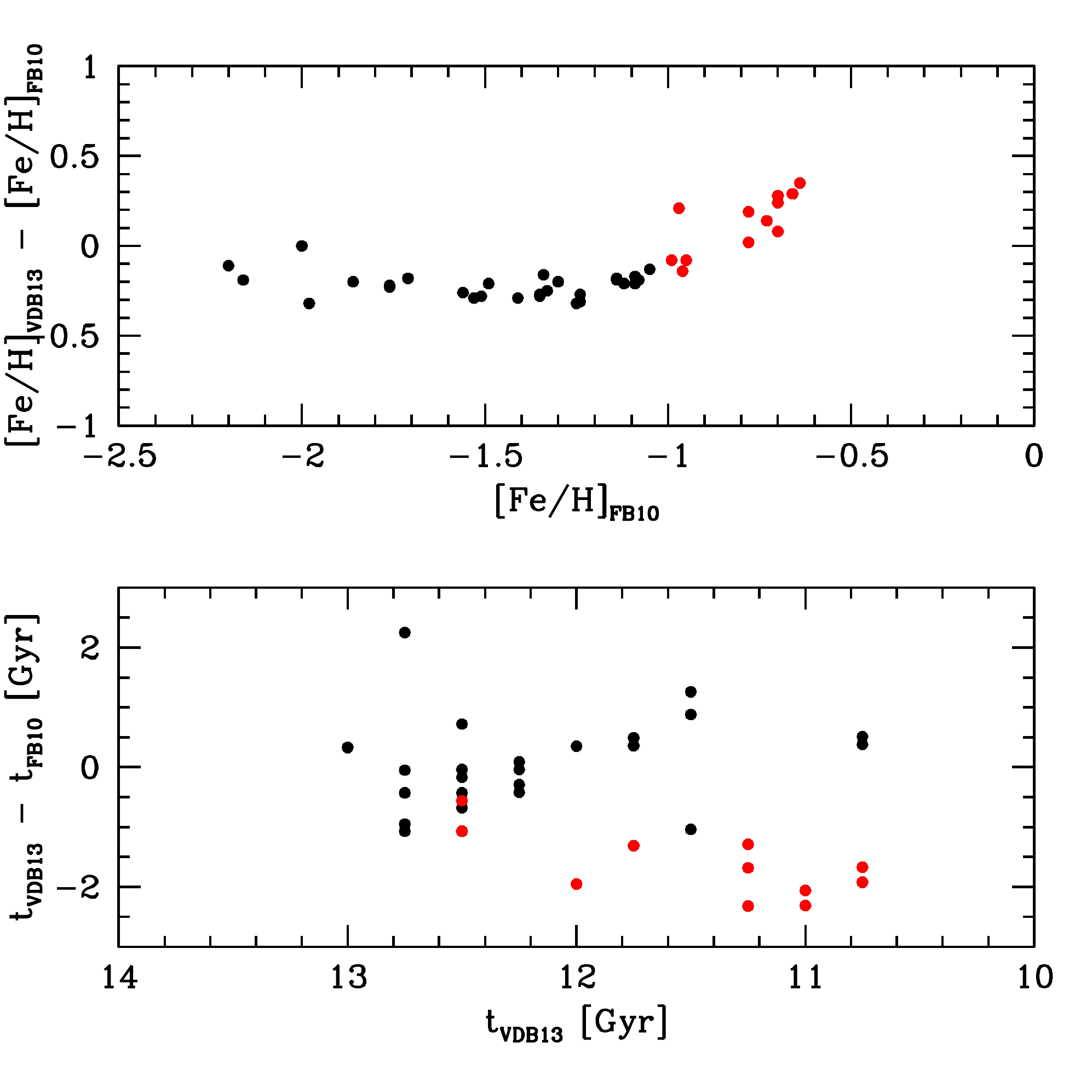}
        \caption{Comparison between the age and metallicity for the GC lists of \cite{vandenberg13} and \cite{forbes10}.}\label{sys}
\end{figure}

Figure~\ref{sys} highlights the need for care when combining different clusters age
datatsets. The top panel shows the difference in metallity between the
\cite{vandenberg13} estimates based on the spectroscopic scale of
\cite{carretta09}, in comparison to those of \cite{forbes10}. This
difference is constant for [Fe/H]$\lesssim-1.1$, above which it
rises steeply with metallicity to about +0.5 dex. The bottom panel of
Fig.~\ref{sys} shows the effect of this trend on age, namely that the
clusters with [Fe/H]~$\gtrsim-1.1$ (red symbols) appear to
be systematically older by $\sim2$ Gyr.  When excluding these
clusters, the mean difference between the age estimates is
$\Delta t$ = 0.08~Gyr, with a spread of $\sigma_{t}$=0.75~Gyr (r.m.s
= 0.14 Gyr), and thus fairly consistent with zero.  Therefore, we
only consider clusters from the \cite{forbes10} sample with [Fe/H]~$\lesssim-1.1$, and assign an uncertainty to their age estimates
(which lack errors) that equals the observed spread around the mean
difference ($\sigma_{t}=0.75$ Gyr).  As for metallicities, we adopted
the scale from \cite{carretta09}.

We also studied the estimates reported in \cite{rosenberg99}, \cite{dotter11}, \cite{roediger14}, but found either poorly constrained values
(with ages as old as 15 Gyr and very large uncertainties), or no new
entries.  Moreover, we decided not to include age estimates performed
on single objects because of the impossibility of having systematic
effects under control. The only exception is NGC5634, 
estimated to be as old as NGC4372 by \cite{bellazzini02}, and for
which we used NGC4372 age estimate by \cite{vandenberg13}.

\end{appendix}

\end{document}